# METALLICITY EFFECTS ON RR LYRAE VARIABLES

Giuseppe Bono

Osservatorio Astronomico di Trieste, Via G.B. Tiepolo 11, 34131 Trieste, Italy;

bono@oat.ts.astro.it

Roberta Incerpi

Dipartimento di Fisica, Univ. di Pisa, Piazza Torricelli 2, 56100 Pisa, Italy;

roberta@astr1pi.difi.unipi.it

and

Marcella Marconi

Dipartimento di Fisica, Univ. di Pisa, Piazza Torricelli 2, 56100 Pisa, Italy;

marcella@astr1pi.difi.unipi.it

# ABSTRACT

Since the seminal paper of Preston (1959) the existence of a metal rich component among the RR Lyrae variable stars belonging to the solar neighborhood has become a well-known observational evidence. In this investigation the effects of heavy element abundance on theoretical observables is studied for both fundamental and first overtone pulsators at fixed mass ($M/M_\odot$ =0.53) and luminosity level (log $L/L_\odot$ =1.51). In order to cover the observed range of RR Lyrae metallicity the nonlinear models have been computed by adopting three different metal contents (Z=0.0001, 0.001, 0.02).
It has been found that the bolometric amplitude of first overtone pulsators decreases as the metallicity increases, whereas for fundamental pulsators the amplitude is positively correlated with the metal abundance. The physical mechanisms which rule the dependence of luminosity amplitude and limit cycle stability on this parameter are discussed in detail and the observative scenario with which the present calculations should be compared is also outlined.

*Subject headings:* stars: evolution – stars: horizontal branch – stars: oscillations – stars: variables: other

## 1. INTRODUCTION

One of the main results provided by Christy (1966) in his pioneering paper on pulsation properties and modal stability of cluster variables was that the metal content does not affect the physical mechanisms (K and $\gamma$ effects) which drive the pulsation instability. This finding was supported by subsequent linear investigations based on Christy's analytical approximation (Iben & Hucra 1971) and on a cubic spline interpolation of Cox & Stewart radiative opacity tables (Tuggle & Iben 1972 and references therein). The outcome was similar in a nonlinear, radiative regime (Carson, Stothers & Vemury 1981). In the framework of two international projects (OP, OPAL) new evaluations of Rosseland mean opacities (Rogers & Iglesias 1992; Seaton et al. 1994) have recently become available. They are based on an improved treatment of atomic physics for metals and have already supplied the opportunity for properly addressing long-standing problems concerning the discrepancy between pulsational and evolutionary masses in double-mode and beat Cepheids (Moskalik, Buchler & Marom 1992) and in cluster variables belonging to Oosterhoff I and Oosterhoff II globular clusters (Cox 1991). As far as population II variables are concerned, it has been found that due to an increase of the iron "bump" ($T = 2 \times 10^5$ K) and to a small decrease of the second helium ionization peak ($T = 5 \times 10^4$ K) the new opacities cause a small but non negligible change of both fundamental (F) and first overtone (FO) periods. This difference implies a change in the period ratio which, through the Petersen diagram, allows an evaluation of pulsational masses which are in good agreement with evolutionary masses.

The above quoted investigations were devoted to RR Lyrae variables belonging to galactic globular clusters, and therefore the metal abundances taken into account ranged generally from Z=0.0001 to roughly Z$\approx$0.006. However, going back to spectroscopic observations provided by Preston (1959) and more recently by Suntzeff, Kraft & Kinman (1994) and Layden (1995), it is well-established that the sample of RR Lyrae variables observed both in the solar neighborhood and in the field is characterized by a metallicity distribution reaching the solar metal content ([Fe/H]$\approx$ 0). Even though several theoretical papers were devoted to the pulsational scenario of cluster variables (Gehmeyr 1993; Bono et al. 1996a and references therein), which have significantly improved our knowledge of the topology of their instability strip, we still lack the analyses of the modal stability and the pulsation characteristics of metal rich RR Lyrae variables. In order to bridge this theoretical gap we have undertaken a large survey of full amplitude nonlinear, nonlocal, and time-dependent convective models of metal rich pulsators. After a brief discussion concerning the method adopted for handling the opacity tables we present the effects of metallicity on pulsational amplitudes and modal stability of RR Lyrae variables.

## 2. NEW OPACITIES AND METALLICITY EFFECTS

The reason why the old analytical approximation of "King Ia" and "King IIa" Los Alamos opacity tables provided by Stellingwerf (1975a, b) has been used so widely by the pulsational fraternity during the last twenty years is that it allows an easy handling of opacity changes inside the envelope structure of variable stars. Moreover, due to its intrinsic smoothness, Stellingwerf's analytical approximation (SAA) allows to compute numerically the opacity derivatives with respect to temperature and volume. This approach presents *inter alia* at least three relevant advantages: 1) since the K mechanism strongly depends on opacity derivatives, the analytical approximation overcomes the inherent difficulties in properly evaluating the derivative discontinuities at the table grid points; 2) the SAA, due to the low numerical complexity, is much less cpu time consuming; 3) different chemical compositions can be simultaneously taken into account. A large number of opacity tables have been recently produced, and therefore, instead of attempting an analytical approximation of these tables, we decided to follow a different root for the evaluation of the opacity and its derivatives. The calculations presented in this investigation have been performed by adopting a fixed helium content (Y=0.28). Since both OP and OPAL tables cover the temperature range $100 \geq logT_6 \geq 0.006$, for temperatures lower than 10,000 K we adopted the molecular opacities provided by Alexander & Ferguson (1994). As a first step we computed new much finer mesh opacity tables by interpolating the original tables by means of bi-cubic splines (Seaton 1993). These new tables present an increment of a factor ten in the number of both temperature and density grid points. The new grid points have been distributed in order to increase the zoning of the original tables in which the opacity shows relevant peaks (i.e hydrogen ionization region (HIR), helium ionization region (HEIR) and the iron "bump"). The use of two-dimensional analytical functions, as already noted by Seaton (1993), implies that the first and second derivatives of the opacity with respect to temperature and density are continuous at each mesh point and therefore we adopted the analytical derivatives of bi-cubic splines instead of estimating them numerically. Finally, for avoiding spurious wiggles the opacity and its derivatives in both linear and nonlinear codes are evaluated by means of a bi-linear interpolation of the finer mesh tables.

Fig. 1 shows the logarithm of opacity and the opacity derivatives versus the logarithm of the temperature for the linear model at $M/M_\odot$ =0.53, log $L/L_\odot$ =1.51 and $T_e$ =6500 K. This model is located approximatively close to the center of the instability strip. As already suggested by Kanbur & Simon (1993), OP and OPAL opacities show only negligible differences which cannot affect the physical structure of variable stars. This notwithstanding, the comparison between new and SAA radiative opacities presents some features that are worth mentioning. Even though it has been already pointed out that the new opacities present a slightly reduced peak in coincidence of HIR and second HEIR, in comparison with the SAA opacities the most relevant discrepancy was ascribed to the "K-bump" caused by bound-bound transitions of iron. This discrepancy is also shown in Fig. 1, but let us point out that the difference, at least for RR Lyrae variables, is less evident than previously claimed (Rogers & Iglesias 1992). This finding is due to two different causes. The former is that the SAA was originally derived for providing a better accuracy along the diagonals of the opacity tables since the envelope structure of variable stars moves along these lines during the pulsation cycle. The latter is that the density of the regions located at $T \geq 10^5$ K is relatively high and therefore the iron bump does not show, for this chemical composition, its usual increase. For the same model Fig. 1 also shows the analytical (OP, OPAL) and numerical (SAA) derivatives of opacity with respect to temperature and volume[1]. A glance to these curves suggests that both analytical and numerical derivatives present smooth changes along the whole structure of the envelope and that the difference is mainly due to differences in the opacity. Indeed, they are almost equal up to temperatures of the order of $10^5$ K, whereas at higher temperatures the analytical derivatives, due to the appearance of the "K-bump", show a different trend in comparison with the numerical ones.

---

[1] $KT = \left(\frac{\partial log K}{\partial log T}\right)_V$ whereas $KV = \left(\frac{\partial log K}{\partial log V}\right)_T$.

On the basis of canonical HB evolutionary tracks (Bono et al. 1996b) we adopted a luminosity level of log $L/L_\odot$ =1.51 and a stellar mass of $M/M_\odot$ =0.53 for properly defining the pulsational properties of metal rich (Z=0.02) RR Lyrae variables. Since we are interested in studying the metallicity effects on the pulsational properties of both F and FO we have selected two models characterized by effective temperatures of 7000 K and 6700 K respectively. The quoted models were computed by adopting three different metal abundances, namely Z=0.0001, Z=0.001 and Z=0.02. Complete details for constructing the pulsation models are given in Bono & Stellingwerf (1994, hereinafter referred to as BS). Table 1 summarizes the results of the full amplitude models computed in this investigation. Concerning the metallicity effects as a first point it is to note that F and FO nonlinear periods strictly increase with metallicity. The top panel of Fig. 2 shows the light curves of the FO models for different assumptions on metal abundance. These curves have been plotted so that the phase of minimum radius takes place at $\phi$ =0.5. The quoted figure displays quite clearly that the amplitudes of FO pulsators, moving from lower to higher metal contents, decreases of roughly 0.2 mag. Moreover, the metallicity increase also causes a remarkable change in the light curve shape. In fact the dip, which appears on the rising branch before the luminosity maximum, becomes much less evident whereas the rising branch becomes less steep. As a consequence the rise time (i.e. the fraction of period spent to cover the phase interval between minimum and maximum in luminosity) increases from 0.24 at Z=0.0001 to 0.32 at Z=0.02.

The effects of metallicity on light curves are tightly connected with the efficiency of convective transport. Even though this model is located close to the FO blue edge and therefore at an effective temperature for which the energy transport should be mainly radiative, the increase in metal abundance causes an increase in the envelope opacity and in turn an increase in the efficiency of convective motions. The increase in convective efficiency as flux carrier involves a general decrease of pulsational amplitudes. In order to show the physical context where this mechanism takes place we plotted in the bottom panel of Fig. 2 the convective timescale of HIR and HEIR. The convective timescale at fixed temperature is defined as CTS=$\lambda/\sqrt{\omega}$ where $\lambda$ is the mixing length and $\sqrt{\omega}$ the convective velocity (see BS for a detailed discussion). The convective transport in the HIR results particularly efficient throughout the pulsation cycle (CTS/P$<<$ 1), only close to the phases of minimum radius it becomes less efficient for metal poor cases (CTS/P$\approx$0.6). The convective timescales referred to the HEIR have been plotted along the full cycle only where it is shorter than the pulsation period. These curves show that at fixed pulsation phase the metal rich case presents more efficient convective motions. This case presents also a longer phase interval ($0.2 \leq \phi \leq 0.75$) during which the timescale of dissipative processes is shorter than the dynamical timescale. We note in passing that the pulsational damping operated by convection also affects the modal stability. Indeed the effective temperature of metal rich FO blue edge is 100 K cooler in comparison with the metal poor cases.

Table 1. Dependence of amplitudes on metallicity.

| First Overtone | | | | | | |
|---|---|---|---|---|---|---|
| Z | $T_e$ | Period | $\Delta M$[a] | $\Delta u$[b] | $\Delta T_e$[c] | RT[d] |
| | K | days | mag. | $kms^{-1}$ | K | |
| 0.0001 | 7300 | .2356 | .680 | 61.15 | 1250 | .38 |
| 0.001 | | .2374 | .592 | 54.43 | 1100 | .33 |
| 0.02 | | .2407 | — | — | — | — |
| 0.0001 | 7000 | .2703 | .636 | 77.96 | 1150 | .24 |
| 0.001 | | .2728 | .582 | 72.02 | 1050 | .25 |
| 0.02 | | .2772 | .454 | 39.78 | 800 | .32 |
| Fundamental | | | | | | |
| 0.0001 | 6600 | .4430 | .929 | 103.92 | 1350 | .14 |
| 0.001 | | .4494 | .987 | 107.73 | 1650 | .12 |
| 0.02 | | .4710 | 1.09 | 95.35 | 1750 | .15 |
| 0.0001 | 5900 | .6576 | — | — | — | — |
| 0.001 | | .6642 | — | — | — | — |
| 0.02 | | .6815 | .279 | 33.41 | 500 | .16 |

[a]Bolometric amplitude. [b]Velocity amplitude.

[c]Temperature amplitude. [d]Rise time.

Fig. 3 shows the same quantities plotted in Fig. 2 but they are referred to F models characterized by the same mass value and luminosity level whereas the effective temperature is 6600 K. The light curves plotted in this figure clearly show that bolometric amplitudes of F pulsators, in contrast with FO ones, are correlated with metal content. This effect is the ultimate consequence of two different causes: 1) the efficiency of convective transport during a full cycle is higher for the metal rich case and hence its pulsational amplitudes present a global decrease in comparison with metal poor pulsators. However, the luminosity amplitude does not follow this trend. This peculiar behavior is due to the fact that the metal rich model is characterized by less efficient dissipative processes in comparison with metal poor cases only during the phases preceding the minimum radius (see Fig. 3). 2) In the metal rich model the "K-bump" connected with iron produces a small but non negligible driving to the pulsation in the regions located close to T=300,000 K and a remarkable reduction of radiative damping (see Fig. 4). As a consequence, this model presents a stronger tendency toward destabilization than the metal poor cases. It is noteworthy that in FO models the increase of the metal content causes a reduction of driving. In fact, the linear nodes are located quite close to the lower boundary of the damping region and therefore these layers cannot reduce the damping since they are at rest throughout the pulsation cycle. Moreover, the driving related to the HEIR, due to the intervening effect of heavy elements, is almost halved in comparison with the metal poor cases. The secondary features (bump, dip) which appear in the light curve present a remarkable dependence on metal abundance. As matter of fact along the decreasing branch the metal poor cases show a bump of at least 0.1 mag., whereas in the metal rich case it is barely noticeable. On the other hand the dip in both F and FO pulsators vanishes more and more as soon as the metallicity increases.

The limit cycle stability of F pulsators at the red edge of the instability strip depends on the coupling between radial and convective motions. Due to the positive correlation between the efficiency of convective transport and the metal abundance, the quenching of radial instability in metal rich models is expected to take place at hotter effective temperatures. However, in this context we found that the F red edge of metal rich pulsators is 100 K cooler than the metal poor ones. This result can be easily explained on the basis of the metallicity effects previously discussed. In short, moving toward the red boundary the convection deepens into the H and He ionization regions and then progressively quenches the radial instability. However, metal rich pulsators, due to the reduction of radiative damping, are quenched at lower effective temperatures.

## 3. CONCLUSIONS

The results presented show that FO amplitudes decrease as the metal abundance increases whereas the F amplitudes present an opposite dependence on this parameter. The metal abundance slightly affects the modal stability of RR Lyrae variables in coincidence of the FO blue edge and F red edge. However, the temperature shifts do not modify the width of the instability strip since at the quoted boundaries they are counter-balanced. Moreover, nonlinear calculations firmly suggest that secondary features and rise time of light curves can be adopted for evaluating the metallicity of RR Lyrae variables -for a different approach based on Fourier parameters see Kovacs & Zsoldos (1995)- as soon as they have been properly calibrated for our particular convective model. New series of full amplitude models at different luminosity levels and with a finer temperature resolution are required before a sound comparison with observations can be provided. Thanks to the substantial advancements in the accuracy and completeness of photometric databases concerning RR Lyrae variables belonging to the Galactic bulge (Udalski et al. 1994), to the solar neighborhood (Layden 1995) and to the central region of the LMC bar (Alcock et al. 1996), we plan to compare the theoretical scenario we are constructing with the quoted RR Lyrae samples. Indeed, this approach is of utmost importance for properly constraining the pulsational and the evolutionary properties of these objects. In particular, as we have already shown for cluster variables (Bono et al. 1996a), the Bailey diagram could prove to be the fundamental tool not only for estimating the possible spread in metallicity but also for evaluating the leading physical parameters (mass, luminosity and temperature) of field RR Lyrae variables.

---

## 4. FIGURE CAPTIONS

**Fig. 1.** - The top panel shows the logarithm of opacity versus the logarithm of temperature for a linear radiative model located at $T_e$=6500 K. The chemical composition has been assumed fixed (Y=0.28, Z=0.02) and the curves are referred to different opacity sets: OP (solid line), OPAL (dashed-dotted line) and SAA (dashed line). The middle and lower panels show the temperature and volume derivatives of opacity.

**Fig. 2.** - *Top:* First overtone bolometric light curves versus phase for three different assumptions on metal contents. The curves have been plotted so that the phase of minimum radius takes place at $\phi = 0.5$. *Bottom:* Ratio between the convective timescales and the pulsation period versus phase. The two arrows display the curves which are referred to the HIR and the HEIR respectively. The wiggles shown by the HIR convective timescale for all Z close to $\phi = 0.8$ are due to sudden shifts of the ionization front.

**Fig. 3.** - Same as Fig. 2 but referred to fundamental models located at $T_e$= 6600 K.

**Fig. 4.** - Nonlinear total work integrals versus the logarithm of the external mass for the fundamental models located at $T_e$= 6600 K, surface at right. Different lines display total work integrals referred to different assumptions on metal abundance. The arrows mark the driving regions connected with HIR, HEIR and the "K-bump" of iron.

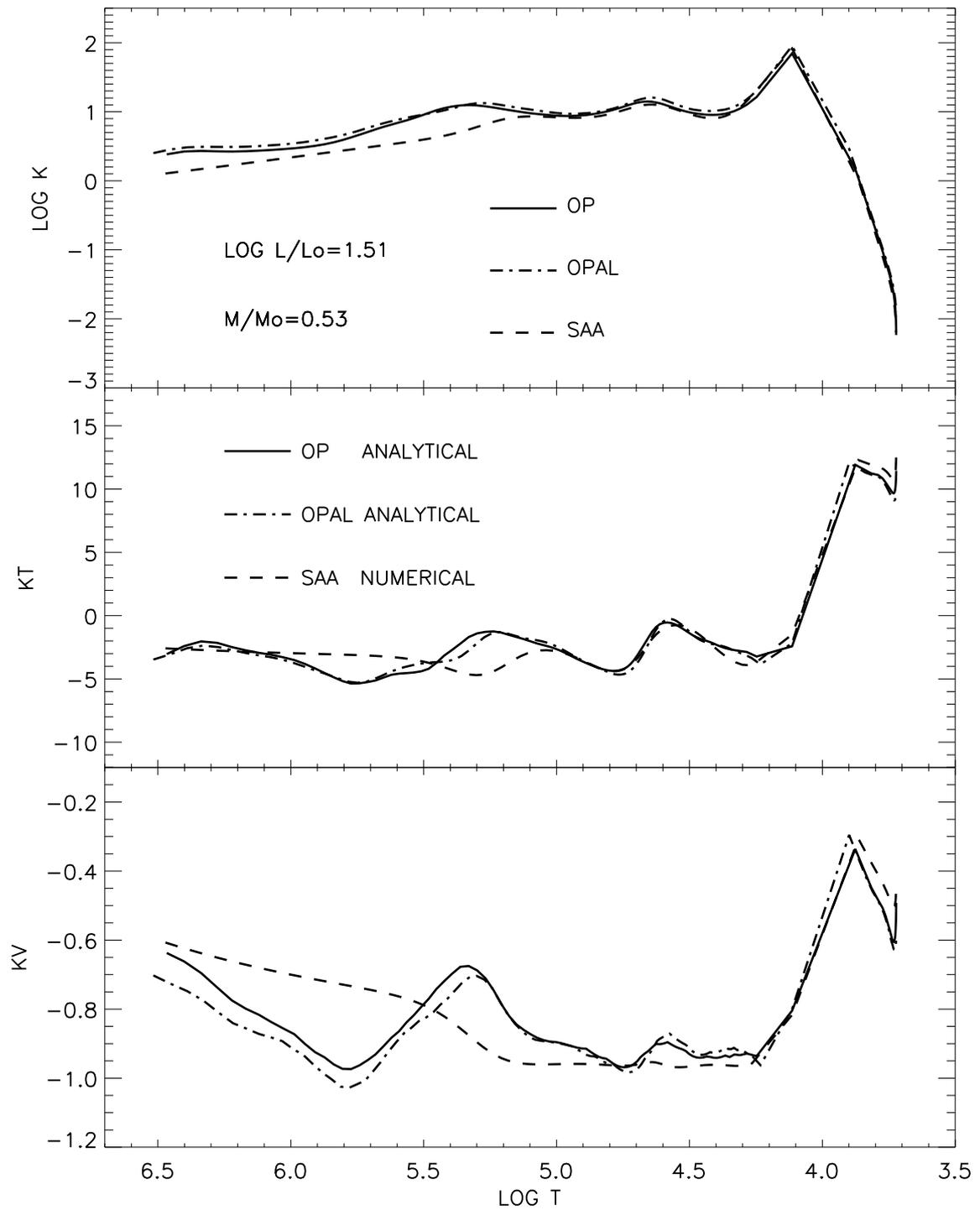

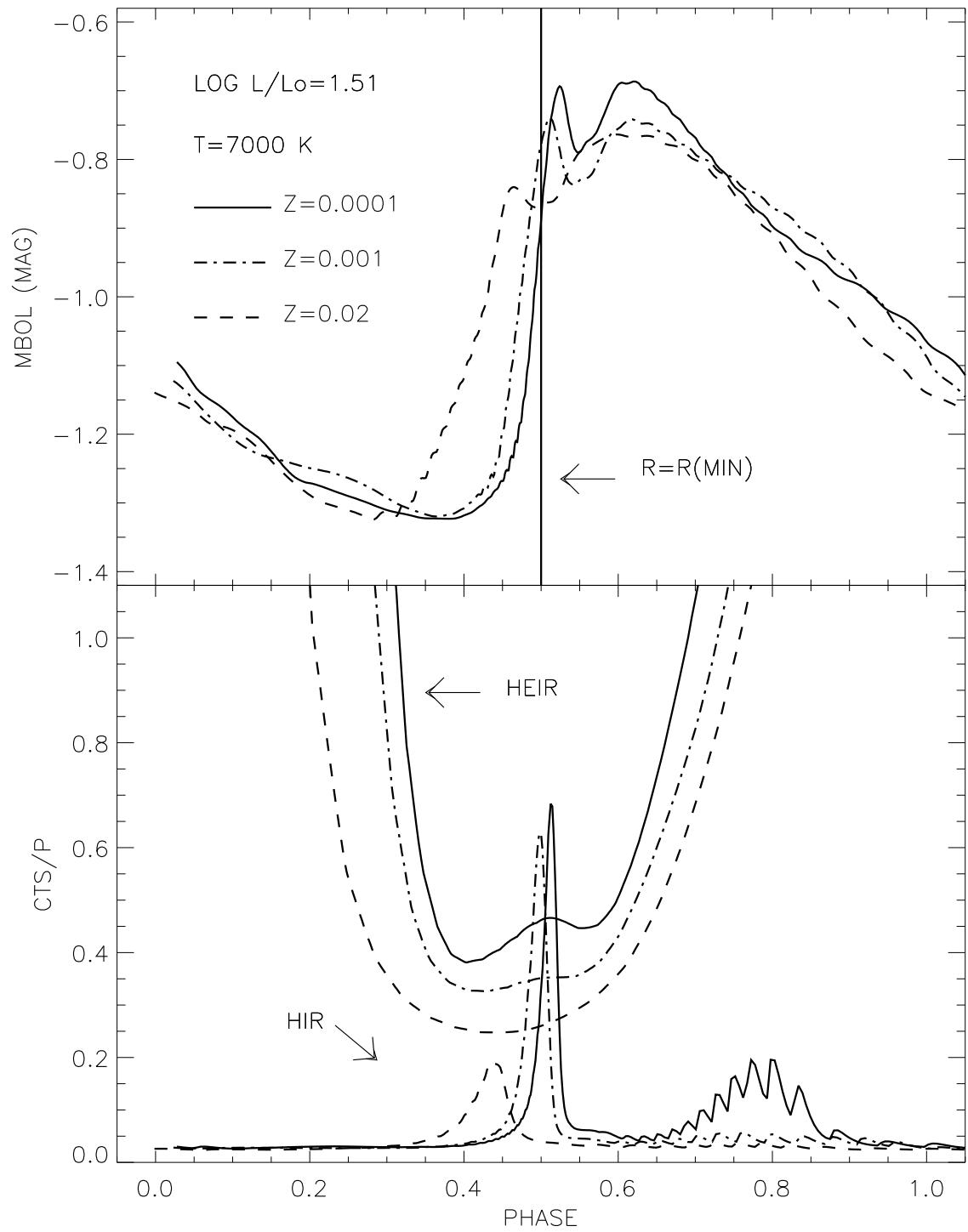

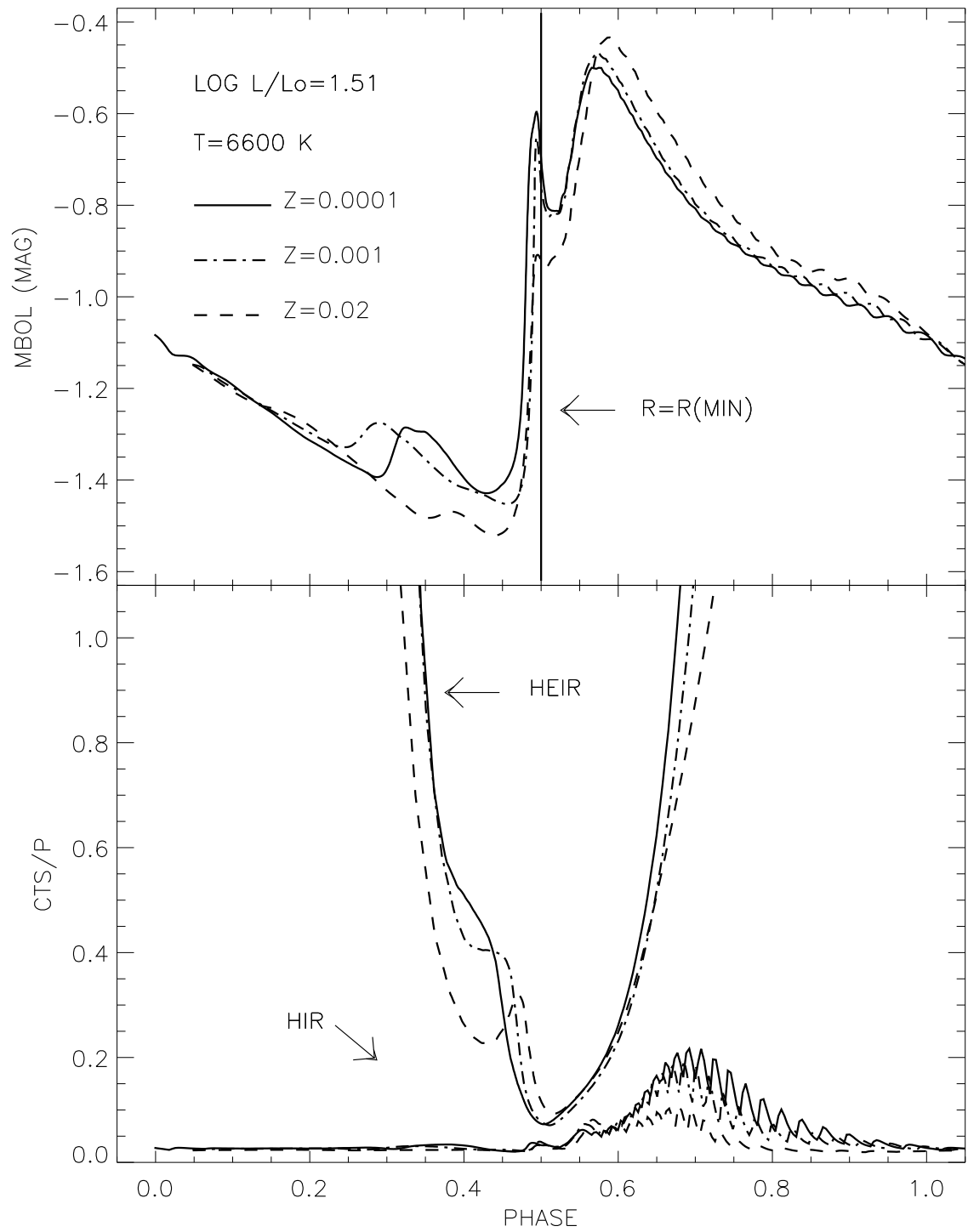

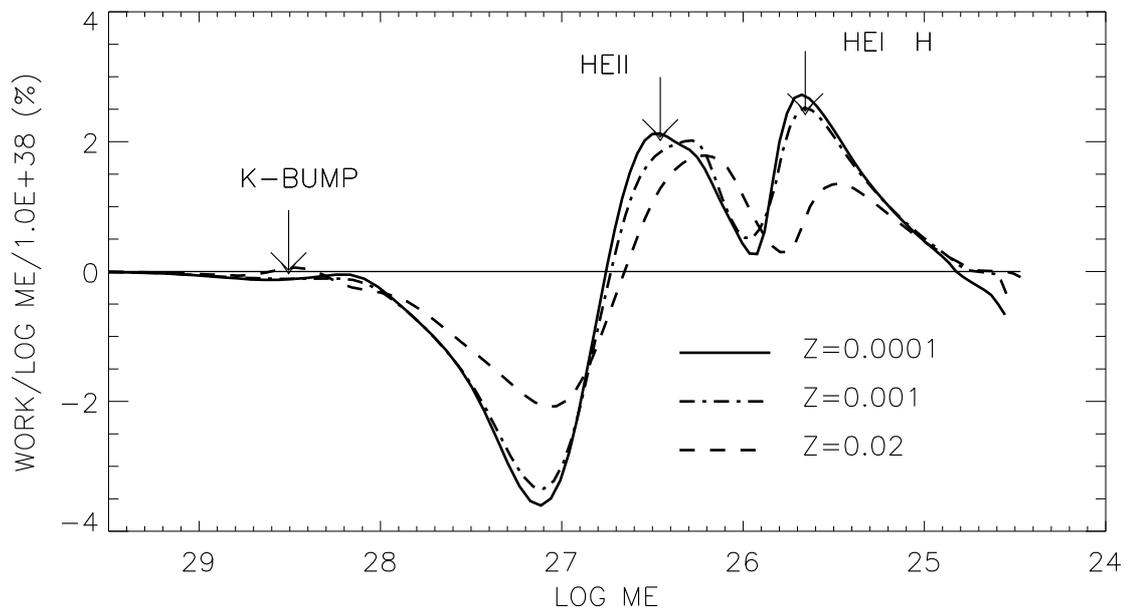